\documentclass[aps,twocolumn,showpacs,preprintnumbers,prb,superscriptaddress]{revtex4-2}

\usepackage[english]{babel}

\usepackage{graphics}
\usepackage{color}
\usepackage{amssymb}
\usepackage{amsmath}
\usepackage{overpic}
\usepackage{epstopdf}
\usepackage{threeparttable}
\usepackage{lipsum}
\usepackage{bm}
\usepackage{braket}
\usepackage{graphicx}
\usepackage{subfigure}
\usepackage{multirow}
\usepackage{array}
\usepackage{newtxtext}
\usepackage[dvipsnames]{xcolor}
\usepackage{glossaries}
\usepackage{siunitx}

\usepackage{hyperref}
\hypersetup{plainpages=false,colorlinks=true,linkcolor=blue, citecolor=blue, urlcolor=blue}

\usepackage{longtable}
\usepackage{booktabs}
\usepackage{blindtext}

\setacronymstyle{long-short}
\newacronym{ft}{FT}{force theorem}
\newacronym{sc}{SC}{self-consistent}
\newacronym{oma}{OMA}{orbital moment anisotropy}

\begin{document}

\title{First-principles prediction of phase transition of YCo$_5$ from self-consistent phonon calculations}

\author{Guangzong Xing}
\email{XING.Guangzong@nims.go.jp}
\affiliation{Research Center for Magnetic and Spintronic Materials, National Institute for Materials Science, Tsukuba, Ibaraki, 305-0047, Japan}

\author{Yoshio Miura}
\affiliation{Research Center for Magnetic and Spintronic Materials, National Institute for Materials Science, Tsukuba, Ibaraki, 305-0047, Japan}

\author{Terumasa Tadano}
\email{TADANO.Terumasa@nims.go.jp}
\affiliation{Research Center for Magnetic and Spintronic Materials, National Institute for Materials Science, Tsukuba, Ibaraki, 305-0047, Japan}

\date{\today}

\begin{abstract}
Recent theoretical study has shown that the hexagonal YCo$_5$ is dynamically unstable and distorts into a stable orthorhombic structure. In this study, we show theoretically that the orthorhombic phase is energetically more stable than the hexagonal phase in the low-temperature region, while the phonon entropy stabilizes the hexagonal phase thermodynamically in the high-temperature region. The orthorhombic-to-hexagonal phase transition temperature is $\sim$165~K, which is determined using the self-consistent phonon calculations. We investigate the magnetocrystalline anisotropy energy (MAE) using the self-consistent and non-self-consistent (force theorem) calculations with the spin-orbit interaction (SOI) along with the Hubbard $U$ correction. Then, we find that the orthorhombic phase has similar MAE, orbital moment, and its anisotropy to the hexagonal phase when the self-consistent calculation with the SOI is performed. Since the orthorhombic phase still gives magnetic properties comparable to the experiments, the orthorhombic distortion is potentially realized in the low-temperature region, which awaits experimental exploration.

\end{abstract}

\maketitle

\section{Introduction}

CaCu$_5$-type RECo$_5$ (RE = rare earth) intermetallic compound has emerged as promising permanent magnets owing to its high Curie temperature, saturation magnetization, and strong coercivity~\cite{RECo5_3,RECo5_4,RECo5_review,patrick.2017.rare,SmCo5}. The coercivity is related to the intrinsic properties of a material, namely magnetocrystalline anisotropy energy (MAE), which can be obtained by measuring the energy difference along different directions of magnetization with respect to the crystal axes~\cite{mca,mca_1,YCo5_mca_exp,YCoFe5_mca_exp1}. Among the RECo$_5$ family, magnetic properties of YCo$_5$ have been extensively studied both from experiments and by first-principles methods based on density functional theory (DFT)~\cite{patrick.2017.rare,YCo5_mca_lapw}. The saturated magnetization of $\sim$910~emu/cm$^3$, the magnetocrystalline anisotropy constant of 7.38$\times$10$^7$~erg/cm$^3$ at 4.2~K, and the Curie temperature of 977~K were obtained in Ref.~\cite{YCo5_mca_exp}. Previous studies indicate that DFT calculations at the level of local density approximation (LDA) or the generalized gradient approximation (GGA) significantly underestimate the orbital magnetic moments of Co atoms and the MAE in comparison with the experimental values. These underestimation problems have been tackled by including the orbital polarization scheme~\cite{YCo5_orbit,YCo5_mag_FLAPW} or by using DFT+$U$~\cite{YCo5_mca_GGAU,mca_self_con} or LDA+DMFT~\cite{Zhu.Batista.2014} approaches. 

\begin{figure*}
    \centering
    \includegraphics[width=0.90\textwidth]{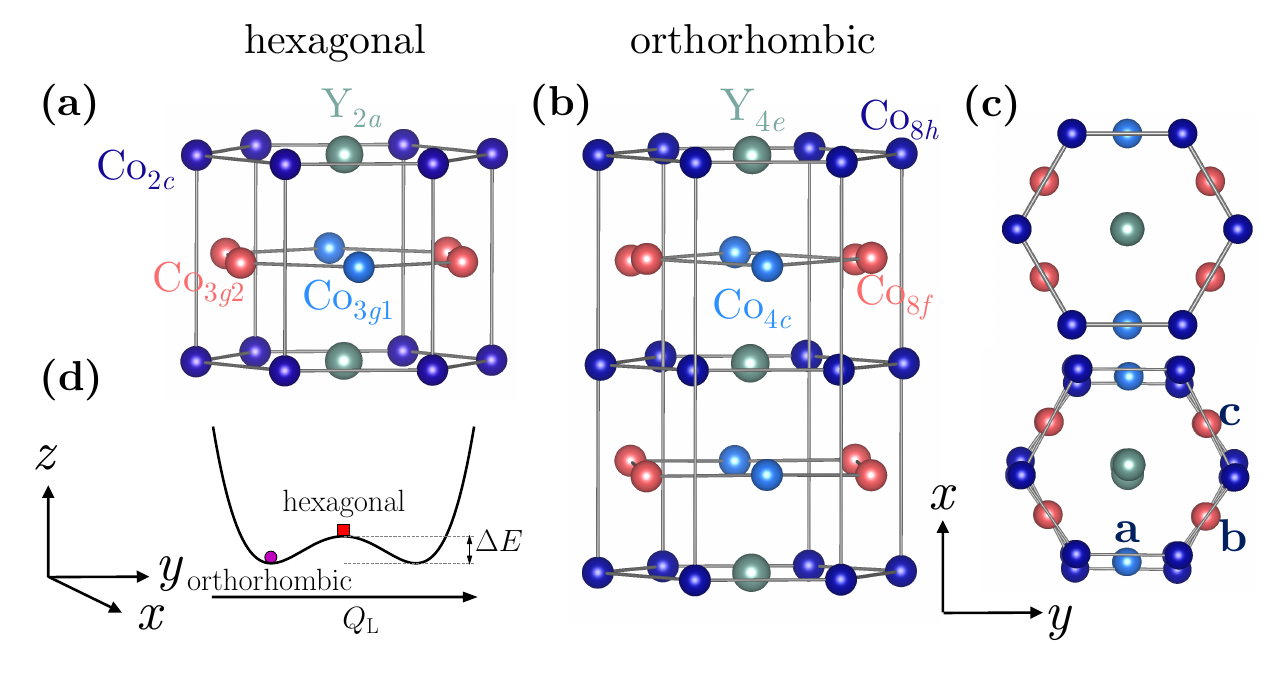}
    \caption{Crystal structure of the hexagonal (a) and orthorhombic (b) phase of YCo$_5$. The Y and Co atoms are represented by the large and small spheres, respectively. The atomic Wyckoff position for different atoms is marked with the corresponding colors. (c) Top view of the crystal structures of hexagonal (top panel) and orthorhombic (lower panel) YCo$_5$. (d) Schematic diagram of the double-well potential obtained by displacing atoms along the direction of the $\mathrm{L}$-point soft phonon mode of the hexagonal phase. $Q_{\mathrm{L}}$ is the normal coordinate amplitude, and } $\Delta{}E$ represents the static energy difference between the two phases.
    \label{fig.1:strcut}
\end{figure*}

All of the previous theoretical calculations have been performed using the CaCu$_5$-type structure [Fig.~\ref{fig.1:strcut}(a)], which displays a layered hexagonal lattice (space group: $P6/mmm$) with two different kinds of Co atoms labeled as Co$_{2c}$ and Co$_{3g}$~\cite{Co3g}. However, a recent first-principles study has shown that a phonon at the L point of the first Brillouin zone is dynamically unstable at 0 K~\cite{YCo5_lattice_ha}, indicating that the hexagonal phase distorts into a lower-energy phase, as schematically illustrated in Fig.~\ref{fig.1:strcut}(d). Indeed, the previous study has reported that the distorted orthorhombic phase (space group: $Imma$) is energetically more stable than the hexagonal phase. Given the presence of the double-well potential energy surface of the unstable phonon mode [Fig.~\ref{fig.1:strcut}(d)], there should be a structural phase transition induced by the soft phonon excited at finite temperatures. Since the magnetic properties, such as MAE, is sensitive to the crystal structure, understanding the temperature evolution of the crystal structure is crucial for reliably comparing the magnetic properties of YCo$_{5}$ obtained from theory and experiments.

Here, by using the state-of-the-art anharmonic phonon calculation method, we demonstrate that the orthorhombic-to-hexagonal phase transition of YCo$_5$ can take place with heating. We first show the orthorhombic phase predicted to be \textit{energetically} more stable than the hexagonal phase at 0 K with various exchange-correlation (XC) functionals, including GGA, GGA+$U$, meta-GGA SCAN, and HSE06. As the temperature rises, the hexagonal phase becomes \textit{thermodynamically} more stable than the orthorhombic phase at $\sim$165~K, which is estimated from the Helmholtz free energies computed using the self-consistent phonon method~\cite{phonons_SCP0,phonons_SCP}. The MAE values of both phases computed using the force theorem, which is a non-self-consistent calculation with the spin-orbit interaction (SOI), within GGA underestimate the experimental values, whereas a self-consistent calculation with the SOI under the GGA$+U$ scheme yields larger MAE values comparable to the experiments. By contrast, the MAE values in the orthorhombic phase obtained by the force theorem within GGA$+U$ still fall short of experimental values for all the applied $U$ values. We attribute this to the fact that both the orbital moment of $\sim$0.1 $\mu_{\mathrm{B}}$ and \gls{oma} with the value of $\sim$0.03 $\mu_{\mathrm{B}}$, which is proportional to the MAE, at the Co site were significantly underestimated by the force theorem compared to the corresponding experimentally measured values of $\sim$0.2 $\mu_{\mathrm{B}}$\cite{Zhu.Batista.2014} and $\sim$0.06 $\mu_{\mathrm{B}}$\cite{YCo5_mca_exp}, respectively. 

The structure of this paper is organized as follows. The theoretical approach to calculate MAE and computational details are described in Secs.~\ref{subsec:MAE_0K} and \ref{subsec:computational_detail}, respectively. The main results are shown in Sec.~\ref{sec:results}. The orthorhombic-to-hexagonal phase transition induced by lattice anharmonicity is discussed in Sec.~\ref{subsec: SCP}. In Sec.~\ref{subsec:MAE}, we evaluate the MAE for both phases using both force theorem and self-consistent methods within the GGA$+U$. Finally, we summarize the study in Sec.~\ref{sec:summary}. 

\section{Methods}

\label{sec:method}

\subsection{MAE from DFT calculation}

\label{subsec:MAE_0K}

The MAE at 0~K was evaluated based on DFT as 
\begin{equation}
    K_{\mathrm{u}}^{\mathrm{DFT}} = E_{0,\perp}-E_{0,\parallel},\label{eq:K_DFT}
\end{equation}
where $E_{0,\perp}$ and $E_{0,\parallel}$ are the ground state total energies with the magnetic moment ($\bm{m}$) being aligned along the hard and easy axes, respectively. These energies were computed by performing DFT calculations with the SOI using the force theorem~\cite{force_theorem} (a non-self-consistent approach) and self-consistent approaches. The easy axis for both phases was found along the [001] direction ($\bm{m}\parallel\bm{c}$). To determine the hard axis, we calculated the total energy by varying the azimuthal angle $\phi$ in the basal plane ($\bm{m}\perp \bm{c}$) from 0 to 2$\pi$ with a step of $\frac{\pi}{6}$. The energy difference between different $\phi$ in the hexagonal phase was negligible. In the orthorhombic phase, the maximum energy difference was $\sim$0.4~MJ/m$^3$ with the lowest energy at $\phi=\frac{\pi}{2}$ ([010] direction). Therefore, we employed the hard axis along the [100] direction for the hexagonal phase and along the [010] direction for the orthorhombic phase to obtain $K_{\mathrm{u}}^{\mathrm{DFT}}$. Although the choice of the hard axis may slightly affect the MAE, the small energy difference does not affect the overall conclusion.

According to the Bruno relation~\cite{Bruno_relation}, MAE is proportional to the \gls{oma} defined as
\begin{equation}
    \Delta{}m_{\mathrm{o}}=m_{\mathrm{o,\parallel}}^{\mathrm{Co}}-m_{\mathrm{o,\perp}}^{\mathrm{Co}},
\end{equation}
where $m_{\mathrm{o,\parallel}}^{\mathrm{Co}}$ and $m_{\mathrm{o,\perp}}^{\mathrm{Co}}$ are the orbital magnetic moment at the Co site along the easy and hard axes, respectively.

\subsection{Computational details}
\label{subsec:computational_detail}

The DFT calculations in this work were performed mainly by using the projector augmented wave (PAW) method~\cite{paw}, as implemented in \textsc{VASP} code (version 6.2.1)~\cite{vasp}. The recommended set of the PAW potentials, which accounts for the scalar relativistic effect, was used. A kinetic-energy cutoff of 400 eV and a $k$-point mesh density of $\sim$450~\si{\angstrom}$^3$ were employed. We adopted the second-order Methfessel-Paxton~\cite{Methfessel_Paxton1989} (MP) smearing method with a width of 0.2 eV for structural optimization. 
A much denser $k$-mesh density of $\sim$6000~\si{\angstrom}$^3$ and the tetrahedron method with the Bl\"{o}chl correction~\cite{Blochl_PRB1994} were employed to calculate the static energy difference $\Delta{}E=E_0^{\mathrm{orth}}-E_0^{\mathrm{hex}}$ between the orthorhombic and hexagonal phase, the MAE, the density-of-state (DOS), and the crystal orbital Hamilton populations (COHP). For these calculations, the initial local moments of 3~$\mu_{\mathrm{B}}$ and \num{-0.3}~$\mu_{\mathrm{B}}$ were used for the Co and Y atoms, respectively. The SOI was also included in the MAE calculation. The subsequent phonon calculations were performed within the GGA by Perdew, Burke, and Ernzerhof (GGA-PBE)~\cite{pbe-gga} without the Hubbard $U$ correction for the following reasons. First, as we elaborate below, we confirmed that the orthorhombic phase is energetically more stable at 0~K, irrespective of the XC functionals. Second, phonon frequencies are sensitive to the employed lattice parameters, and GGA-PBE yields lattice parameters of the hexagonal phase that are in accord with the experimental values~\cite{YCo5_thermal} within the error of $\sim$1\%.  Lastly, we encountered technical challenges in achieving convergences when applying GGA$+U$ functionals for displaced supercells. The calculated potential energy surface was not smooth enough to obtain reliable harmonic force constants, despite repeated adjustments to the mixing parameters eventually led to convergence.

It has been reported that MAE obtained using GGA-PBE significantly underestimated the experimental values. To address the underestimation problem, we employ the GGA$+U$ scheme~\cite{LDAU} to account for the correlation effect of the Co-$3d$ electrons. To compute the total energies under the GGA$+U$ scheme with the SOI, we employed two different approaches: \gls{sc} and the \gls{ft} calculations. The main difference is whether the charge density from the self-consistent spin-polarized calculation is updated self-consistently or not; in the SC method, the charge density is optimized again, whereas it is not so in the FT. Additionally, we employed the full-potential linearized augmented plane-wave (FLAPW) method~\cite{singh.2006.planewaves}, as implemented in the WIEN2k code~\cite{schwarz.2002.electronic}, to investigate the effect of the core electrons on the MAE and orbital moment of YCo$_5$. The crystal structure optimized using VASP within GGA-PBE was adopted to calculate the MAE, the OMA, and the orbital moment. We used LAPW sphere radii of 2.015 and 2.115 for Co and Y, respectively, and a basis set cut-off parameter of $R_{\mathrm{min}}K_{\mathrm{max}}=9$.

\begin{table}[!b]
\centering
  \caption{Calculated the static electronic energy difference ($\Delta{}E$, see text) between the hexagonal and orthorhombic phases and the squared frequency of the soft mode at the L point ($\omega_{\mathrm{L}}^2$) for hexagonal YCo$_5$ using different XC functionals.}
  \label{table:deltaE}
  \small
  \begin{ruledtabular}
  \begin{tabular}{lcc}
  XC-functional & $\Delta{}E$ (meV/f.u.) & $\omega_{\mathrm{L}}^2$ (cm$^{-2}$) \\
    \toprule[0.15mm]
    LDA     & $-44.08$ & $-7826.4$ \\
    PBE & $-23.24$ & $-4301.2$\\
    PBEsol & $-23.78$ & $-4425.9$ \\
    GGA
+$U$ ($U=1$ eV) & $-52.39$ & $\cdots$\\ 
    SCAN  & $-97.31$ & $\cdots$ \\
    HSE06 & $-134.48$ & $\cdots$ \\
  \end{tabular}
  \end{ruledtabular}
\end{table}

To compute the vibrational Helmholtz free energy $F_{\mathrm{vib}}$ within the first-order self-consistent phonon (SCP) theory, second-order and fourth-order interatomic force constants (IFCs) are required. A $2\times2\times2$ supercell containing 48 (96) atoms was used for calculating the second-order IFCs of the hexagonal (orthorhombic) phase. Here, each atom in the supercell was displaced from its equilibrium position by 0.03~\si{\angstrom}, and the atomic forces were calculated using VASP. From the generated displacement-force dataset, we estimated the second-order IFCs by ordinary least squares. For anharmonic IFCs, a supercell containing 48 atoms was adopted for both phases. We employed the compressive sensing lattice dynamics to extract the anharmonic IFCs~\cite{compre_learm} from the displacement-force training datasets. Here, we uniformly sampled 200 training structures from the last \num{2500} steps of an \textit{ab initio} molecular dynamics calculation (\num{5000} steps in total) at 10~K for both phases. For each sampled structure, we further displaced all atoms by 0.1~\si{\angstrom} in random directions. We have confirmed that 200 sample structures were sufficient and the change in $\Delta{F_{\mathrm{vib}}}=F_{\mathrm{vib}}^{\mathrm{orth}}-F_{\mathrm{vib}}^{\mathrm{hex}}$ between the orthorhombic and hexagonal phase after adding 50 more structures (250 structures in total) was smaller than $\sim$0.5 meV/f.u. at 500~K.  In our phonon calculations based on GGA-PBE, we found that the harmonic phonon dispersion of the hexagonal phase is sensitive to the broadening parameter $\sigma$ of the MP method used for the IFCs calculations, which is particularly noticeable when $\sigma\gtrsim 0.1$ eV. Thus, we carefully investigated the $\sigma$ dependence of the predicted structural transition temperature ($T_\mathrm{c})$ and found that $\sigma$ as small as 0.085 eV was desirable for obtaining a converged $T_c$ value. We discuss this point in Sec. \ref{subsec: SCP}.
All the phonon calculations in this study were performed using the $\textsc{alamode}$ code~\cite{ALAMODE,phonons_SCP}.

\section{Results and Discussion}
\label{sec:results}

\subsection{Phase transition induced by lattice anharmonicity}
\label{subsec: SCP}

First, to assess the stability of the hexagonal and orthorhombic phases at 0~K, we calculated the ground state energies using various XC functionals: LDA~\cite{kohn.1965.lda}, PBE, its variant for solids (PBEsol)~\cite{perdew.2008.pbesol}, GGA+$U$~\cite{LDAU} with $U=1$ eV, the strongly constrained and appropriately normed (SCAN) meta-GGA~\cite{sun.2015.scan}, and the Heyd-Scuseria-Ernzerhof (HSE)06 hybrid functional~\cite{krukau.2006.hse06}. For each XC functional other than HSE06, the crystal structures of the two phases were fully relaxed with the collinear ferromagnetic spin configuration. The detailed structural parameters are listed in Tables~S1 and S2 of the supplemental material (SM)~\cite{supplement}. For HSE06, the crystal structures from the PBE calculation were used. The difference of the ground state energy $\Delta E=E_0^{\mathrm{orth}}-E_0^{\mathrm{hex}}$, where $E_0^{X}$ being the energy of the phase $X$ per formula unit (f.u.), is summarized in Table \ref{table:deltaE}. It is clear that the orthorhombic phase is predicted to be energetically more stable at 0 K irrespective of the employed XC functionals. In the following, the crystal structures for both phases optimized by GGA-PBE  are used to perform further phonon and magnetic property calculations except for the squared frequency $\omega_{\mathrm{L}}^2$ of the soft-mode tabulated in Table~\ref{table:deltaE}, for which the crystal structures optimized by each XC functional are used.

\begin{figure}
    \centering
    \includegraphics[width=0.45\textwidth]{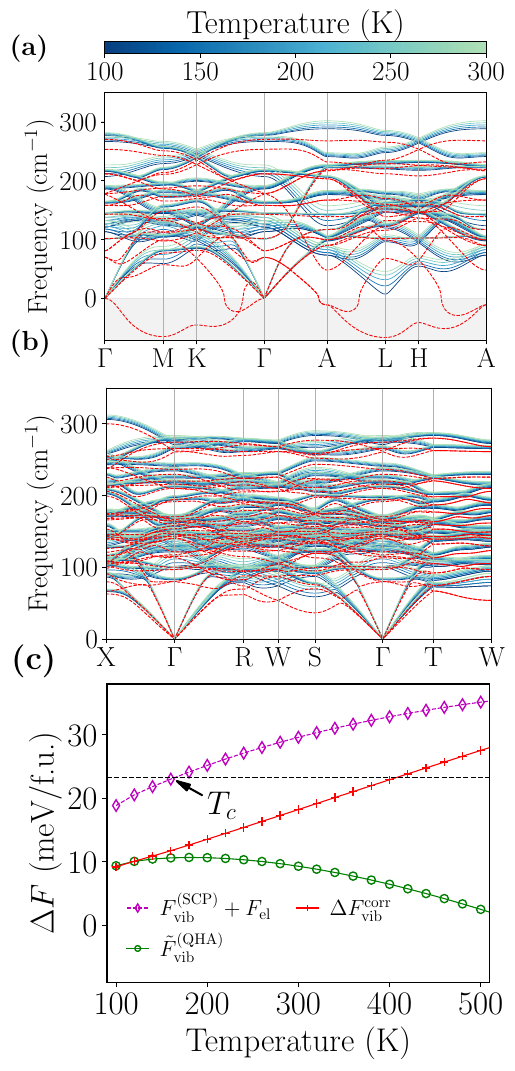}
    \caption{Temperature-dependent anharmonic phonon dispersion of hexagonal (a) and orthorhombic (b) phases of YCo$_5$. The colormap shows the self-consistent phonon solutions at the temperature range of 100--300~K, and the dashed lines are harmonic lattice dynamics results. Imaginary frequency is shown as negative. (c) Calculated difference of the free energies, $\Delta{}F(T)$ (see text). The horizontal dashed line is the difference of static energy $-\Delta{E}=23.24$ meV/f.u.  Here, the phonon calculation are performed with $\sigma=0.075$ eV.}.
    \label{fig.2:phonon}
\end{figure}

Next, we discuss the thermodynamic stability of the two phases at finite temperatures by comparing the Helmholtz free energies defined as
\begin{align}
    F(V,T) = E_0(V_0)+F_{\mathrm{vib}}(V,T)+F_{\mathrm{el}}(V,T), \label{eq:free_energy}
\end{align}
where $E_0(V)$ is the static electronic energy obtained from a conventional DFT calculation, and $F_{\mathrm{el}}(V,T)$ and $F_{\mathrm{vib}}(V,T)$ are the electronic and vibrational free energy at temperature $T$ and cell volume $V$. Since the lattice constants reportedly increase only by $\sim$0.04 \AA{}  at 600 K~\cite{YCo5_thermal}, we neglect the thermal expansion effect and use the optimized cell volume $V_0$ at 0~K for computing the Helmholtz free energy. Also, since the magnetic phase transition, e.g., anomaly in the heat capacity, has never been observed in the low-temperature region, the magnetic entropy is also omitted in this study assuming that its difference between the two phases is negligible. Then, the electronic free energy $F_{\mathrm{el}}(V_0, T)$ and vibrational free energy $F_{\mathrm{vib}}(V_0,T)$ play a central role in determining the thermodynamic stability. $F_{\mathrm{el}}$ is obtained based on the fixed density-of-states approximation~\cite{xing.2021.lattice,togo.2015.first}. When all phonon modes are dynamically stable within the harmonic approximation (HA), it is straightforward to estimate $F_{\mathrm{vib}}(V_0,T)$. However, for the hexagonal YCo$_{5}$, the method based on the HA breaks down due to the presence of the soft mode at the L point. Hence, in this study, we employ the SCP scheme. In this approach, we first compute finite-temperature phonon frequencies by solving the following SCP equation~\cite{phonons_SCP4,phonons_SCP5}
\begin{align}
    \Omega_q^2=\omega_q^2+\frac{1}{2}\sum_{q_1}\Phi(q;-q;q_1;-q_1)\alpha_{q_1}. \label{eq:scp}
\end{align}
Here, $\omega_{q}$ is the angular frequency of the phonon mode $q$ obtained within the HA, $\Omega_{q}$ is the frequency at finite temperature renormalized by the anharmonic interaction $\Phi(q;-q;q_1;-q_1)$ associated with the fourth-order anharmonicity of the potential energy surface, and $\alpha_{q}=\frac{\hbar}{2\Omega_{q}}[1+2n(\Omega_{q})]$ with $n(\omega)$ being the Bose-Einstein distribution function. Once the above equation is solved for $\Omega_{q}$, the vibrational free energy can be evaluated as $F_{\mathrm{vib}}^{\mathrm{(SCP)}}(V_0,T)= \tilde{F}_{\mathrm{vib}}^{\mathrm{(QHA)}}(V_0,T)+\Delta{}F_{\mathrm{vib}}^{\mathrm{corr}}(V_0,T)$, which is the sum of the quasi-harmonic (QH) term computed using $\Omega_{q}$ and the correction term necessary to satisfy the correct thermodynamic relationship of $-dF_{\mathrm{vib}}/dT=S_{\mathrm{vib}}$~\cite{F_vib_SCP,F_vib_SCP1}. More details of the mathematical expressions for $F_{\mathrm{vib}}^{\mathrm{(SCP)}}$ can be found in the SM~\cite{supplement}.

\begin{figure}[t]
    \centering
    \includegraphics[width=0.45\textwidth]{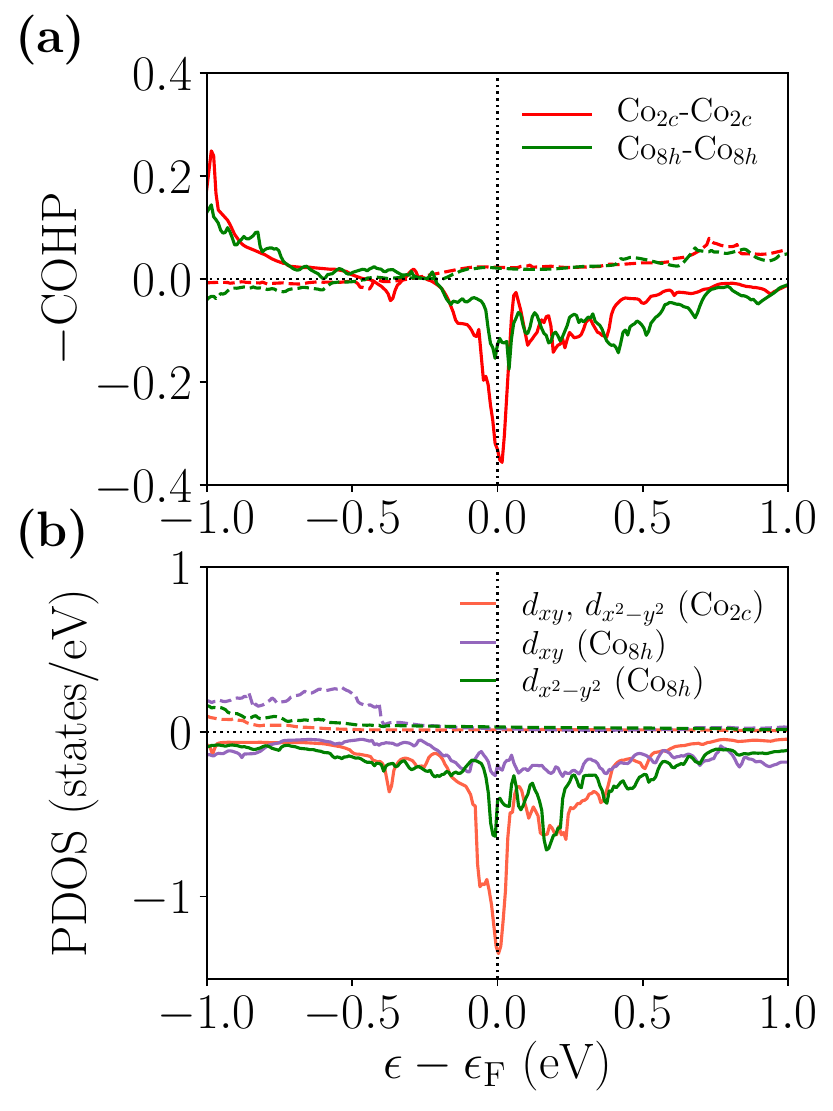}
    \caption{(a) Crystal orbital Hamilton populations (COHP) calculation of the Co$_{2c}$-Co$_{2c}$ bond in the hexagonal phase and Co$_{8h}$-Co$_{8h}$ bond in the orthorhombic phase. (b) Projected density of states (DOS) for $3d$ states at Co$_{2c}$(Co$_{8h}$) site. The solid and dashed lines represent the minority and majority spin states, respectively. The data of the hexagonal YCo$_5$ is adapted from Ref.~\cite{YCo5_lattice_ha}.} 
    \label{fig.3:cohp}
\end{figure}

Figure~\ref{fig.2:phonon}(a) and (b) show the phonon dispersion curves of hexagonal and orthorhombic YCo$_5$ within the HA (dashed lines) and SCP theory (solid lines). The SCP frequencies from 100~K to 300~K in the step of 50~K are shown with different colors. It is clear that the harmonic phonon of the hexagonal YCo$_5$ is dynamically unstable with the most unstable mode occurring at the L points. This instability was also observed with other XC functionals, as indicated by the negative $\omega_\mathrm{L}^2$ values in Table~\ref{table:deltaE}. By contrast, the harmonic phonon is dynamically stable for the distorted orthorhombic phase [see Fig.~\ref{fig.2:phonon}(b)]. The soft modes of the hexagonal phase at L and M points mainly involve the in-plane [$xy$-plane in Fig.~\ref{fig.1:strcut}(c)] displacements of the Co$_{2c}$ atoms together with relatively small displacements of the other atoms. Recently, it has been shown that the phonon instability at the L point originates from the strong antibonding nature of the Co$_{2c}$-Co$_{2c}$ bond~\cite{YCo5_lattice_ha}, as evidenced by a sharp peak in the crystal orbital Hamilton populations (COHP)~\cite{COHP} near the Fermi level. To see how the antibonding nature changes with the structural distortion, we compare the calculated $-$COHP values in Fig.~\ref{fig.3:cohp}(a). The originally equidistant Co$_{2c}$-Co$_{2c}$ bonds split into three different  Co$_{8h}$-Co$_{8h}$ bonds labeled as $a$, $b$, and $c$ [see bottom panel of Fig.~\ref{fig.1:strcut}(c)], with the corresponding bond length of 2.75~\AA,{} 2.57~\AA, and 3.21~\AA{}, respectively. Hence, the $-$COHP values for the three inequivalent Co$_{8h}$-Co$_{8h}$ bonds are averaged here. Compared with the large population of antibonding states [red curve in Fig.~\ref{fig.3:cohp}(a)] at Fermi energy in hexagonal YCo$_5$, the magnitude of $-$COHP in the orthorhombic phase is reduced significantly, which can be explained by the smaller projected DOS [Fig.~\ref{fig.3:cohp}(b)] of the $d_{x^2-y^2}$ orbital at the Co$_{8h}$ sites. Such a large reduction in the $-$COHP value by structural distortion is consistent with the trend of phonon stability we observed. Interestingly, the projected DOS in Fig.~S1 of the SM~\cite{supplement} shows that the peak of the Co$_{2c}$ $d_{x^2-y^2}$ orbital at the Fermi energy does not change appreciably even when we used the SCAN functional and GGA$+U$ with various $U$ parameters. This suggests the presence of phonon instability in the hexagonal YCo$_5$ even with the XC functionals other than GGA-PBE, in accord with the trend of $\Delta E$ shown in Table \ref{table:deltaE}.

As can be seen in Figs.~\ref{fig.2:phonon}(a) and \ref{fig.2:phonon}(b), the phonon frequencies significantly increase by the quartic anharmonicity at finite temperatures, particularly notable in the low-energy optical modes. Since SCP theory postulates the renormalized phonons to be stable in Eq.~(\ref{eq:scp}), we observe that the finite-temperature phonons are stable irrespective of the temperature for both phases. Hence, the information of the self-consistent phonon dispersion alone cannot determine the structural phase transition temperature $T_c$ reliably~\cite{SCP_static,SCP_appr}. In this study, we employ another approach based on Eq.~(\ref{eq:free_energy}), where $T_c$ is estimated from the difference in the electronic and vibrational free energy defined as 
$\Delta{}F(T)=F_{\mathrm{vib}}^{\mathrm{orth}}(T)+F_{\mathrm{el}}^{\mathrm{orth}}(T)-[F_{\mathrm{vib}}^{\mathrm{hex}}(T)+F_{\mathrm{el}}^{\mathrm{hex}}(T)]$. The calculated $\Delta{}F(T)$ is shown in Fig.~\ref{fig.2:phonon}(c). The increase of $\Delta{}F$ with heating occurs because the electronic and vibrational entropy gradually enhances the stability of the hexagonal phase. Eventually, the hexagonal phase becomes more stable when $F^{\mathrm{hex}}(T)\leq F^{\mathrm{orth}}(T)$, or equivalently $-\Delta E \leq \Delta F(T)$, is satisfied. From this condition, $T_c$ of $\sim$165 K was obtained when evaluating $\Delta{}F(T)$ using the MP method with the broadening parameter $\sigma$ of 0.075~eV, as shown in Fig.~\ref{fig.2:phonon}(c). While a magneto-elastic study showed that the hexagonal phase to be stable down to $\sim$100~K under hydrostatic pressure~\cite{rosner.2006.magneto,koudela.2008.magnetic}, a detailed structure analysis at ambient pressure below $\sim$165 K is necessary to test the present prediction.

Here, we discuss the dependency of $T_\mathrm{c}$ on the employed $\sigma$ value. As shown in Fig.~S2 of the SM~\cite{supplement}, the harmonic phonon dispersion of the hexagonal phase was sensitive to the $\sigma$ value,  particularly noticeable near the A $(0,0,\frac{1}{2})$ point of the Brillouin zone. By contrast, we did not observe such a significant $\sigma$ dependence for the phonon dispersion of the orthorhombic phase. These different behaviors are consistent with the large difference in the projected DOS at the Fermi level, shown in Fig.~\ref{fig.3:cohp}(b). Interestingly, in the case of anharmonic phonon calculations, the phonon frequencies were not so sensitive to $\sigma$ even for the hexagonal phase. This occurred presumably because the relatively large and random displacements we used for computing anharmonic IFCs lifted the degeneracy and thereby reduced the DOS at the Fermi level, making the anharmonic IFCs less sensitive to $\sigma$. Because of the observed $\sigma$ dependency of the phonon frequencies, the $\Delta F(T)$ value also changed slightly with $\sigma$, leading to a sizeable shift in $T_\mathrm{c}$, as shown in Fig.~S3(b) of the SM~\cite{supplement}. With decreasing $\sigma$ from 0.2 eV to 0.085 eV, $T_{\mathrm{c}}$ gradually decreased from $\sim$240 K to $\sim$165 K. In the smaller $\sigma$ region of 0.060--0.085 eV, the calculated $T_{\mathrm{c}}$ value remained nearly constant at $\sim$165 K, indicating convergence of the calculation with respect to $\sigma$. We also confirmed that a $\sigma$ value as small as 0.085 eV can give a potential energy surface of the third mode at the A point which agrees almost perfectly with that computed using the tetrahedron method with the Bl\"{o}chl correction, whereas a larger $\sigma$ overestimated the Hessian, as shown in Fig.~S2(b) of the SM~\cite{supplement}.

\begin{figure*}
    \centering
    \includegraphics[width=0.98\textwidth]{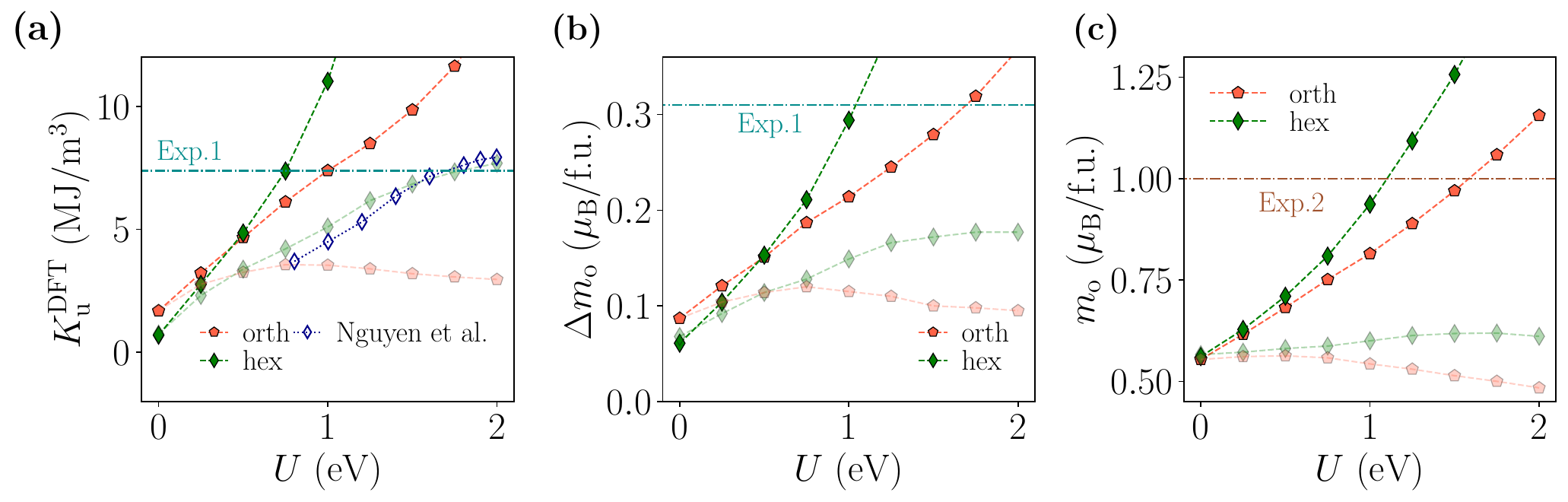}
    \caption{ Dependence of magnetocrystalline anisotropy energy, $K_{\mathrm{u}}^{\mathrm{DFT}}$, (a) and orbital moment anisotropy, $\Delta{m_{\mathrm{o}}}$, (b) and orbital moment (c) for the hexagonal and orthorhombic phase of  YCo$_{5}$ under the GGA$+U$ scheme. The solid and translucent symbols represent the quantities mentioned above obtained using the self-consistent and force theorem approaches, respectively. The unfilled diamond is the theoretical results from Ref.~\cite{YCo5_mca_GGAU} based on the force theorem, and the horizontal dash-dotted lines, labeled as Exp.1, and Exp.2 are experimental results from Ref.~\cite{YCo5_mca_exp}, and Ref.~\cite{Zhu.Batista.2014}, respectively. Note that the $\Delta{m_{\mathrm{o}}}$ and $m_{\mathrm{o}}$ of Y atom are not included in (b) and (c) for comparison with the experimental data.} 
    \label{fig.4:K_u}
\end{figure*}

\subsection{MAE for two phases}
\label{subsec:MAE}
In the following, we compare the MAE values of the two phases obtained from two different approaches: \gls{ft}~\cite{force_theorem}, i.e., a non-self-consistent calculation with the SOI, and \gls{sc} calculations with the SOI. For the hexagonal phase, the MAE value, defined as $K_\mathrm{u}^{\mathrm{DFT}}=E_{0,\perp}-E_{0,\parallel}$, was 0.75 MJ/m$^3$ within GGA-PBE, which significantly underestimates the experimental values of $\sim$7.4~MJ/m$^3$~\cite{YCo5_mca_exp} and $\sim$6.0~MJ/m$^3$~\cite{YCoFe5_mca_exp1}. This underestimation problem could be resolved by adding the Hubbard $U$ correction. Zhou \textit{et al.}~\cite{zhou.2009.obtaining} reported that the GGA$+U$ method does not adequately describe the orbital ground state for strongly correlated $f$-electron systems due to the significant anisotropy in the self-interaction error (SIE) of the $f$ orbitals.  However, in YCo$_5$ without $f$ electrons, the 3$d$ electrons of Co are more itinerant and less correlated, and the SIE anisotropy should be less significant in comparison with the typical energy scale of the crystal field and bandwidth~\cite{richter.1998.band,YCo5_orbit}. Therefore, the GGA$+U$ method is still valid for calculating the MAE and orbital moment, which has been demonstrated in the previous GGA$+U$ studies on some Co compounds~\cite{donati.2013.magnetic,tung.2012.anomalous}.

For hexagonal YCo$_5$,  our \gls{ft} calculation in Fig.~\ref{fig.4:K_u}(a) nicely reproduced the $U$ dependence of MAE from a previous \gls{ft} result based on GGA$+U$~\cite{YCo5_mca_GGAU}, and the $K_{\mathrm{u}}^{\mathrm{DFT}}$ value became comparable with the experimental values when $U\sim$1.75 eV. However, at this $U$ value, the \gls{ft} still underestimated the total orbital moments and their anisotropy (\gls{oma}), $\Delta{m_{\mathrm{o}}}=m_{\mathrm{o,\parallel}}^{\mathrm{Co}}-m_{\mathrm{o,\perp}}^{\mathrm{Co}}$, for all Co sites, as shown in Figs.~\ref{fig.4:K_u}(c) and (b), respectively. We observed that the \gls{sc} approach can yield the MAE and orbital moment results much more consistent with the experimental values, as we elaborate below. The \gls{sc} scheme enhanced the $K_{\mathrm{u}}^{\mathrm{DFT}}$ values at all the $U$ values, as shown in Fig.~\ref{fig.4:K_u}(a). Consequently, an agreement with experimental values was obtained for MAE at $U\sim$0.75 eV, which is smaller than the $U$ of $\sim$1.75 eV for the \gls{ft} approach. At $U=$0.75 eV, the \gls{sc} approach gave the orbital moment of $\sim$0.8 $\mu_{\mathrm{B}}$/f.u., in better agreements with the experimental value of $\sim$1 $\mu_{\mathrm{B}}$/f.u.~\cite{Zhu.Batista.2014} [see Fig.~\ref{fig.4:K_u}(c)]. Although perfect agreements in the MAE, orbital moment, and \gls{oma} were not reached simultaneously with a single $U$ value, it is clear that the \gls{sc} scheme gives more reliable results than the \gls{ft} approach.

For the orthorhombic phase, we obtained the $U$-dependent MAE, \gls{oma}, and the orbital moment which were qualitatively similar to those of the hexagonal phase when the \gls{sc} approach was used, as shown by filled markers in Figs.~\ref{fig.4:K_u}(a), (b), and (c). One minor difference can be found in the $U$ value to reproduce experimental MAE; for the orthorhombic phase, a slightly larger $U$ value of $\sim$1.0 eV appears to give essentially the same magnetic properties as the hexagonal phase. Even at $U=0.75$ eV, which was applied to the hexagonal phase, the MAE of 6.12~MJ/m$^3$ can still reach the range of experimental observations (6.0-7.4~MJ/m$^3$). Interestingly, we found that the \gls{ft} failed completely for the orthorhombic phase, as inferred from the consistently lower MAE values in Fig.~\ref{fig.4:K_u}(a), which can be attributed to the underestimation of the \gls{oma} [Fig.~\ref{fig.4:K_u}(b)]. In the context of the GGA$+U$ scheme with the SOI, the \gls{ft} approach only modifies the GGA energy due to the Coulomb interaction between the Co $3d$ electrons occupying the same ion. However, the change in occupancy due to the on-site electron-electron Coulomb interaction is neglected because the charge density is not self-consistently updated. Therefore, it is reasonable that the FT calculations underestimate the orbital moment, the \gls{oma}, and consequently the MAE. We note that \gls{sc} approach could reproduce the experimentally-observed larger orbital moments and \gls{oma} at the Co$_{2c}$(Co$_{8h}$) than the other Co sites for both of the two phases. The details of the site dependency are summarized in the SM~\cite{supplement}. In addition, the total spin and orbital magnetic moment obtained by the \gls{sc} calculation with $U=0.75$ eV ($U=1.0$ eV) for the hexagonal (orthorhombic) phase is 8.44 $\mu_{\mathrm{B}}$/f.u. (8.52 $\mu_{\mathrm{B}}$/f.u.), and agrees well with the experimental value of $\sim$8.3 $\mu_{\mathrm{B}}$/f.u.~\cite{YCo5_mca_exp} and $\sim$8.4 $\mu_{\mathrm{B}}$/f.u.~\cite{patrick.2017.rare}, while the total moment of 7.80 $\mu_{\mathrm{B}}$/f.u. (7.73 $\mu_{\mathrm{B}}$/f.u.) determined without $U$ underestimates the experimental results.
We have confirmed that other GGA$+U$ correction schemes~\cite{liechtenstein.1995.density,czyzyk.1994.local,ylvisaker.2009.anisotropy} as well as a full-potential all-electron calculation~\cite{singh.2006.planewaves,schwarz.2002.electronic} give quantitatively similar MAE values. (Please see details in the SM~\cite{supplement}).

\section{Summary}

\label{sec:summary}

To summarize, we have theoretically demonstrated a possible orthorhombic-to-hexagonal structural phase transition of YCo$_5$ induced by heating. We found the orthorhombic phase is energetically more stable than the hexagonal phase at 0~K irrespective of the employed exchange-correlation functionals. The stable, temperature-dependent phonons of these phases were obtained by incorporating the anharmonic renormalization using the self-consistent phonon approach, and the theoretical phase transition temperature of $\sim$165~K was obtained by comparing the calculated Helmholtz free energies. We compared the MAE, the associated \acrlong{oma}, and the orbital moments of the two phases computed using the \acrlong{ft} and \acrlong{sc} calculations with the SOI based on the GGA$+U$ scheme. We showed that the magnetic properties of the orthorhombic phase were similar to those of the hexagonal phase when the \acrlong{sc} approach was employed. Since the calculated magnetic properties also agreed well with the available experimental data, we expect the predicted orthorhombic phase to be existent in the low-temperature region, which awaits experimental verification. 

\begin{acknowledgments}

This work was partly supported by MEXT Program: Data Creation and Utilization-Type Material Research and Development Project (Digital Transformation Initiative Center for Magnetic Materials) Grant Number JPMXP1122715503. The figures of crystal structures are created by the \textsc{vesta} software~\cite{vesta}.
\end{acknowledgments}

\bibliography{YCo5_Imma_new.bib}

\begin{thebibliography}{58}
\bibitem [{Co3()}]{Co3g}%
  \BibitemOpen
  \href@noop {} {}\bibinfo {note} {The threefold degenerate {Co$_{3g}$} atoms
  will be separated into two inequivalent sites with multiplicity of 1 (denoted
  as {Co$_{3g1}$}) and 2 (denoted as {Co$_{3g2}$}) when the magnetization is
  along the hard axis ([100] direction).}\BibitemShut {Stop}%
\end{thebibliography}

\end{document}